\documentclass{emulateapj}
\usepackage{color}
\usepackage{multirow}
\usepackage{units}
\usepackage[colorlinks,linkcolor=blue,anchorcolor=green,citecolor=blue]{hyperref}
\shorttitle{GRB 201216C}
\shortauthors{Huang}
\begin{document}
%\title{A time dependent numerical model associated with distant gamma-ray bursts: application to GRB 201216C}
\title{Time dependent numerical model for the very high energy emissions of distant gamma-ray busrt GRB 201216C}
%\author{Yan Huang}
%\affil{School of Physics and optoelectronics engineering, Anhui University, Hefei, 230601, China; hyan623@ahu.edu.cn}

\author{Yan Huang\altaffilmark{1}}
\affil{$^1$ School of Physics and optoelectronics engineering, Anhui University, Hefei, 230601, China; hyan623@ahu.edu.cn\\
}

\begin{abstract}

Recently, the MAGIC Collaboration reported a $\sim 5\sigma$ statistical significance of the very-high-energy (VHE) emission from a distant GRB, GRB 201216C. Such distant GRB may be effectively absorbed by the extragalactic background light (EBL). The origin of the VHE emission from such distant objects is still unknown. 
Here, we propose a numerical model for studying the afterglow emission of this distant GRB. The model solves the continuity equation governing the temporal evolution of electron distribution, and the broad-band observed data can be explained by the synchrotron plus synchrotron self-Compton (SSC) radiation of the forward shock. The predicted observed 0.1 TeV flux can reach $\sim 10^{-9} -10^{-10}\rm erg ~ cm^{-2} ~ s^{-1}$ at $t \sim 10^3 -10^4 \rm s$, even with strong EBL absorption, such strong Sub-TeV emissions still can be observed by MAGIC telescope. Using this numerical model, the shock parameters in the modeling are similar with two other Sub-TeV GRBs (i.e., GRB 190114C and GRB 180720B), implying that the Sub-TeV GRBs have some commonalities: they have energetic burst energy, low circum-burst medium density and low magnetic equipartition factor. We regard GRB 201216C as a typical GRB, and estimate the maximum redshift of GRB that can be detected by MAGIC telescope, i.e., $z \sim 1.6$. We also find that the VHE photon energy of such distant GRB can only reach $\sim 0.1 ~\rm TeV$. Improving the low energy sensitivity of the VHE telescope is very important to detect the Sub-TeV emissions of these distant GRBs.

\end{abstract}


\section{Introduction}

The origin of the very-high-energy (VHE) emissions of Gamma-Ray Bursts (GRBs) has been an open issue for a long time. There are several mechanisms can produce the VHE gamma-ray during the afterglow phase\citep{zhang2019, magic2019b}: (1) Synchrotron radiation from the accelerated electrons in the forward shock. But the radiation has a maximum energy due to the Lorentz factor of the outflow, i.e., $\varepsilon_{\rm max} \sim 100 \rm MeV \times \Gamma_{b}(t)/(1+z)$. The Lorentz factor of GRB outflow $\Gamma_{b}(t)$ is usually on the order of ten during the afterglow phase, then the maximum energy of synchrotron radiation is only about $\sim \rm GeV$; (2) Synchrotron radiation from the accelerated protons in the forward shock. But the protons are much less efficient emitters than the electrons and the requirements to reproduce the observed TeV flux and spectrum are more severe; (3) Inverse Compton (IC) emission by the accelerated electrons of the forward shock, where the seed photons may be X-ray flares\citep{murase2011, zhang2020, zhang2021}, shock breakout emission\citep{wang2006}, hypernova envelope emission\citep{he2009} or the synchrotron photons (which are emitted by the same accelerated electrons)\citep{sari2001}, et al,. 

Recently, the VHE gamma-rays from some GRBs (e.g., GRB 190114C, GRB 180720B and GRB 190829A) have been detected by the Ground-based Imaging Atmospheric Cherenkov Telescope\citep{magic2019a, magic2019b, abd2019, hess2021}, such as the Major Atmospheric Gamma Imaging Cherenkov (MAGIC) telescope, and the High Energy Stereoscopic System (H.E.S.S). The observed VHE photons of these GRBs are more likely to be explained by the IC mechanism of the forward shock in the afterglow phase\citep{wang2019, zhang2019, zhang2020}. 

GRB 201216C is a distant GRB which was newly announced to be detected VHE emission by MAGIC telescope. GRB 201216C is particularly bright and hard, with its GBM light curve shows a broad, structured peak. The duration ($T_{90}$) is about 29.9s (50-300keV), the time-averaged spectrum from $T_0$ - 0.03s to $T_0$+49.665s ($T_0$ is the trigger time of GBM) is best fitted by a band function with $E_{\rm peak} = 326 \pm 7 \rm keV$, $\alpha=-1.06 \pm 0.01$, $\beta=-2.25 \pm 0.03$, and the event fluence (10-1000keV) in this interval is $(1.41 \pm 0.06) \times 10^{-4} \rm erg ~cm^{-2}$\citep{mala2020}. VLT X-shooter identifies the redshift of this GRB is $z=1.1$, making this object is the most distant known VHE sources\citep{vie2020}. Using the Fermi-GBM parameters, the isotropic energy release of this GRB is $E_{\rm iso}=(4.71 \pm 0.16) \times 10^{53} \rm erg$. 

%From the grz photometry (Izzo et al. GCN 29066), they measure a spectral slope $\beta_{\rm opt}=4.1 \pm 0.2$ (assume $F_{\rm \nu} \propto \nu^{\rm \beta_{\rm opt}}$), which is an unusually red value, suggesting a significant extinction. This is confirmed by the optical-to-X-ray spectral index, $\beta_{\rm ox} \sim 0.1$, which indicates a very low optical/X-ray flux ratio, making this a bona fide dark GRB. 

The MAGIC telescope was designed to perform gamma-ray astronomy in the energy range from 50 GeV to greater than 50 TeV. The MAGIC telescope observed GRB 201216C following the trigger by Swift-BAT and Fermi-GBM under good conditions about 57s. The preliminary offline analyses showed an excess above $5\sigma$, compatible with the GRB position\citep{blan2020}. Because the started observing time of this VHE emission is two times larger than the duration of prompt emission, we prefer to believe that the VHE emission originates from the afterglow phase. 

The GRB position was not in the  Fermi-LAT field of view until $T_0$+3500s, and no significant high-energy gamma-ray emission associated with this GRB in the time interval from $T_0$+3500s to $T_0$+5500s. Assume a photon index of $-2$, the corresponding upper limit for energy flux is $3 \times 10^{-10} \rm erg~cm^{-2}~s^{-1}$ ($95\%$ confidence level, 100MeV - 1GeV)\citep{biss2020}. Swift-XRT began to observe from 3 ks to 1938 ks after the BAT trigger. The light curve can be modelled with an initial power-law decay with its decay index of $\alpha_{X}=2.09^{+0.16}_{-0.1}$, followed by a break at $T_0+9078 \rm s$ and the decay index after the break is $\alpha_{X}=1.07^{+0.15}_{-0.1}$. The spectrum can be fitted with an absorbed power-law with a photon spectral index of $\sim 2.35$\citep{evans2021}.

At 177s after the BAT trigger, the Liverpool Telescope (LT) observed the field of swift GRB 201216C and confirmed the optical counterpart for this GRBs. They measured $r=18.38$, the r-band data was calibrated with respect to nearby APASS secondary standard stars\citep{shre2020}. The r-band light curve made along with VLT data point and inferred data from FRAM-ORM show a power law decay in flux vs time with $\alpha_{r} \sim 1.07$\citep{jeli2020}. From the light curve, LT observations seem to be around the peak of afterglow.

We are interested in the origin of VHE emissions from such distant GRB. In this work, we base on the available broad-band data of GRB 201216C and propose a paradigm to explain the VHE emission of GRB 201216C. In \S 2, we perform a numerical model of the afterglow emissions. We consider the details of the dynamics of shock evolution, apply the time-dependent treatment to calculate the temporal electron energy distributions (ED), then derive the synchrotron and synchrotron self-Compton (SSC) emissions. The Klein-Nishina (KN) suppression and the absorption are also considered in this work. In \S 3, we present the details of numerical appoach, and compare the numerical results with the analytical results to test the validity of our code. Our numerical model is applied to the afterglow of GRB 201216C in  \S 4. Finally, we give the discussions and conclusions in \S 5. 

%Note that we use the cgs units, and adopt $Q_x=Q/10^x$ throughout all this paper.

\section{Model description}

\subsection{Hydrodynamic evolution}

For the hydrodynamic evolution of GRB external shock, we refer to \citet{Huang1999}. We consider an impulsive outflow with kinetic energy $E_k$, and the initial Lorentz factor $\Gamma_0$, propagating into an external medium of constant density $n$, typical for the interstellar medium (ISM). The differential dynamic equation of the afterglow shock can be derived as
\begin{equation}
\frac{d \Gamma}{dm} = - \frac{\Gamma^2 -1}{m_{\rm ej} +2\Gamma m},
\label{dyn1}
\end{equation}
which describes the overall evolution of the shocked shell from relativistic phase to non-relativistic phase, where $m_{\rm ej}=E_k/ \Gamma_0 c^2$ is the mass of the outflow, $m=(4 \pi /3) R^3 n m_p$ is the mass of the swept-up external matter, and $\Gamma$ is the bulk Lorentz factor of the shocked shell. In order to obtain the time dependence of $\Gamma$, we make use of
\begin{equation}
\frac{d m}{dR}=4 \pi R^2 n m_{p},
\label{dyn2}
\end{equation}
and 
\begin{equation}
\frac{d R}{dt} = \beta c \Gamma (\Gamma+\sqrt{\Gamma^2 -1}),
\label{dyn3}
\end{equation}
where $t$ is the observer time, $R$ and $\beta=\sqrt{1-\Gamma^{-2}}$ are the radius and velocity of the shell, respectively.

Assume $10\%$ of the kinetic energy is converted into the isotropic energy of prompt emission, then the kinetic energy of the outflow is about $E_k \sim 5 \times 10^{54} \rm erg$. For the ultra-relativistic outflow, the deceleration timescale of the shock is
\begin{equation}
t_{\rm dec}=\left(\frac{17E_k}{1024 \pi n m_{p} c^5 \Gamma_0^8}\right)^{1/3},
\label{t_dec1}
\end{equation}
then we have $t_{\rm dec} \approx 235 \rm s ~E_{\rm k, 54}^{1/3} \Gamma_{0,2}^{-8/3} n^{-1/3}$, where $E_k=10^{54} E_{k, 54} \rm erg$, $\Gamma_0=10^2 \Gamma_{0,2}$ and $n$ is normalized to $1 \rm cm^{-3}$. The observed peak time of r-band light curve is around $\sim 177 \rm s$, very close to this deceleration timescale.

For the non-relativistic outflow, the deceleration timescale of the shock is on the order of years\citep{nakar2011}, much larger than $100 \rm s$ and seriously inconsistent with the peak position of the r-band light curve. Hence, we only consider the ultra-relativistic outflow. Then the comoving time $t'$ can be related to the observer's time $t$ for the ultra-relativistic shock by
\begin{equation}
dt=(1+z) \Gamma (1-\beta) dt' \approx \frac{1+z}{2\Gamma} dt',
\label{t_co}
\end{equation}
and the position of the jet head is $dR=\beta c \Gamma dt'$. Hereafter, the superscript prime ($'$) is used to denote the quantities in the comoving frame of the shock fluid.

\subsection{The energy distribution of shock-accelerated electrons}

The energy distribution of the shock-accelerated electrons behind the shock is usually obtained by solving the continuity equation of nonthermal electrons and it is best to work in the co-moving frame of the shock fluid when we solve the continuity equation. Firstly, we assume that the distribution of injected electrons is taken as a power-law between $\gamma_m'$ and $\gamma_{\rm max}'$($\gamma_{\rm max}'>\gamma_{m}'$),
\begin{equation}
d \dot N_{\rm e,0}'=\dot N_{\rm e}' \frac{p-1}{\gamma_{\rm m}'} \left(\frac{\gamma_{e}'}{\gamma_{\rm m}'}\right)^{-p} d \gamma_{e}',
\end{equation}
with
\begin{equation}
\dot N_{e}'=\Gamma \left(\beta +\frac{1}{3}\right)4 \pi n R^2 c,
\end{equation}
is the number of electrons swept and accelerated by the shock wave per unit time. $\gamma_{e}'$ is the random Lorentz factor of an electron, $\gamma_{m}'$ is the minimum Lorentz factor which can be determined by the shock jump conditions\citep{bla76, sari1998},
\begin{equation}
\gamma_{m}'=\frac{p-2}{p-1} \epsilon_{e} (\Gamma-1) \frac{m_{p}}{m_{e}},
\label{gamma_m}
\end{equation}
and $\gamma_{\rm max}'$ is the maximum Lorentz factor which can be determined by comparing the acceleration time scale $t_{\rm acc}'$ with the cooling time scale $t_{\rm cool}'$. The acceleration time scale is to be $t_{\rm acc}' \approx k_B \gamma'_{e} m_e c/(e B')$, where $k_B$ is the correction factor, which accounts for the downstream electrons do not return to the downstream region once they are deflected by an angle of $1/ \Gamma$, therefore $k_B$ should not be smaller than 1, i.e., $k_B \gtrsim 1$. We take $k_B = 1$ here. The cooling timescale due to synchrotron and IC cooling both is estimated to be $t_{\rm cool}'=(6 \pi m_{e} c/(\gamma'_{e} \sigma_{\rm T} B'^2)) (1+Y)$, where $\sigma_{\rm T}$ is the Thompson scattering cross section, and $Y$ is the Compton parameter that is defined as the ratio of the IC radiation power to the synchrotron radiation power, i.e., $Y= P'_{\rm IC}/P'_{\rm syn}$. Then $\gamma_{\rm max}'$ can be determined by
\begin{equation}
\gamma_{\rm max}'=\sqrt{\frac{6 \pi e}{k_B \sigma_T B' (1+Y)}}.
\label{gamma_max}
\end{equation}
$B'$ is the comoving magnetic field strength. We assume that the magnetic field density behind the shock is a constant fraction $\epsilon_B$, then
\begin{equation}
B'=(32 \pi m_p \epsilon_B n)^{1/2} \Gamma c.
\label{dyn2}
\end{equation}

The afterglow emission is produced by a distribution of relativistic electrons that injected into the downstream of the shock, at a rate $Q(\gamma'_e)=d \dot N_{\rm e,0}'/d \gamma_{e}'$. Here, we treat the shocked region as a thin and uniform shell, the continuity equation governing the temporal evolution of the ED $dN_{e}'/d\gamma_{e}'$ is
\begin{equation}
\frac{\partial}{\partial t'} \frac{dN_{e}'}{d\gamma_{e}'} +\frac{\partial}{\partial \gamma_{\rm e }'} \left(\dot \gamma_{\rm e, tot}' \frac{dN_{e}'}{d\gamma_{e}'}  \right) = Q(\gamma'_e),
\label{continuity equation}
\end{equation}
where $\dot \gamma_{\rm e,tot}'$ is the total cooling rate of an electron, which can be obtained by summing up to the processes of adiabatic cooling $\dot \gamma_{\rm e,adi}'$, synchrotron cooling $\dot \gamma_{\rm e,syn}'$ and IC cooling $\dot \gamma_{\rm e, ic}$,
\begin{equation}
\dot \gamma_{\rm e,tot}'= \dot \gamma_{\rm e,adi}' +\dot \gamma_{\rm e,syn}'+\dot \gamma_{\rm e, ic}'.
\label{gamma_dot}
\end{equation}
The electrons heating rate due to synchrotron self-absorption (SSA) and the electron-positron pair injection due to $\gamma \gamma$-absorption are not considered in Eq.(\ref{continuity equation}). 

\subsection{The cooling and radiative processes}

The adiabatic cooling of the shock is due to the spreading of the shock fluid, which is described by
\begin{equation}
\dot{\gamma}'_{\rm e, adi}=\frac{1}{3} \gamma'_{e} \frac{d \ln n'_e}{dt'}=-\frac{1}{2} \frac{\gamma'_e}{R} \frac{dR}{dt'},
\label{adi}
\end{equation}
where $n'_e \propto R^{-3/2}$ is the comoving electron number density.

The GRB shocks can accelerate the swept-up electrons, compress and amplify the ambient magnetic field, thus the electrons in the shock are losing energy by synchrotron radiation. The synchrotron cooling rate of an electron with Lorentz factor $\gamma_{e}'$ is
\begin{equation}
\dot \gamma'_{\rm e, syn}=- \frac{\sigma_{T} B'^2 \gamma_{e}'^2}{6 \pi m_{e} c}.
\end{equation}

The accelerated electrons are also cooled by IC scattering of the seed photons, such as synchrotron photons, black-body photons from the supernova, and the X-ray flares photons, et al,. The IC emissivity calculated by the Thomson cross section is inaccurate for high-energy gamma-ray photons, so we should use the full KN cross section to instead. The IC cooling rate (only consider the first-order IC component) with considering a significant KN correction is given by\citep{blu70, fan08,geng18},
\begin{equation}
\dot{\gamma}'_{\rm e, ic}=-\frac{1}{m_{e}c^2} \frac{3 \sigma_{T} c}{4 \gamma_{e}'^2} \int^{\nu_{\rm max}'}_{\nu_{\rm min}'} \frac{n_{\nu'} d\nu'}{\nu'} \int^{\nu_{\rm ic, max}'}_{\nu_{\rm ic, min}'} h \nu_{\rm ic}' d \nu_{\rm ic}' F(q,g),
\label{ssc}
\end{equation}
where $F(q,g)=2q \ln q+(1+2q)(1-q)+\frac{1}{2} \frac{(4qg)^2}{1+4gq}(1-q)$, $g=\frac{\gamma_{e}' h \nu'}{m_{e} c^2}$,  $w=\frac{h \nu_{\rm ic}'}{\gamma_{e}' m_{e} c^2}$ and $q=\frac{w}{4g(1-w)}$. $\nu'$ and $\nu'_{\rm ic}$ in Eq.(\ref{ssc}) are the frequencies of the seed photons and the IC photons respectively. According to the dynamics of the collision between a relativistic electron and a photon, i.e., $1 \ll h \nu'/ \gamma_{e}' m_{e} c^2 \leq h \nu_{\rm ic}/ \gamma_{e}' m_{e} c^2 \leq 4g/(1+4g)$, and $1/4 \gamma_{e}' \leq q \leq 1$, the upper limit of the second intergral can be derived as $h \nu_{\rm ic,max}'=\gamma_{e}' m_{e} c^2 \frac{4g}{4g+1}$, and the lower limit is $\nu_{\rm ic,min}'=\nu'$.

GRB 201216C was not accompanied by the obvious X-ray flare, shock breakout and hypernova envelope emission, hence it has been proposed that the VHE $\gamma$-ray photons may be produced by the SSC scenario, then the quantity $n_{\nu'}$ need in Eq.(\ref{ssc}) is the synchrotron seed photon spectra, which is calculated by
\begin{equation}
n_{\rm \nu'} \approx \frac{T'}{h \nu'} \frac{\sqrt{3} e^3 B'}{m_{e} c^2} \int^{\gamma_{\rm e,max}'}_{\gamma_{\rm e, min}'} n'(\gamma_{e}') \mathcal{R} (\nu'/\nu_{c}') d\gamma_e',
\label{n_nv1}
\end{equation}
where $T' \approx \Delta' /c$ is the co-moving time that the synchrotron photons stay within the shell, $n'(\gamma_{e}')=\frac{dN_{e}'/d\gamma_{e}'}{4 \pi \Delta' R^2}$ is the co-moving electron number density, $\Delta'$ is the co-moving width of the shock. One should notice that $\Delta'$ does not appear in Eq.(\ref{n_nv1}), since $\Delta'$ in $T'$ and $n'(\gamma_e')$ are cancelled out. $\nu_c'=3eB' \gamma_e'^2/ 4\pi m_{e} c$ is the critical frequency, \citep{cru86} derived an exact expression for $\mathcal{R}$ in terms of Whittaker's function, $\mathcal{R}$ also can be presented in a simple analytical form\citep{zir07},
\begin{equation}
\mathcal{R} (\nu'/\nu_{c}')=\frac{1.81\exp (-\nu'/\nu_{c}')}{\sqrt{(\nu'/\nu_{c}')^{-2/3}+(3.62/\pi)^2}}.
\end{equation}

After obtaining the solved-out temporal evolution of the ED $dN_{e}' / d\gamma_{e}'$, the total optically thin synchrotron radiation power at frequency $\nu'$ in the comoving frame is\citep{cru86},
\begin{equation}
P_{\rm syn}'(\nu')=\frac{\sqrt{3} e^3 B'}{m_{e} c^2} \int d\gamma_e' \frac{dN_{e}'}{d\gamma_{e}'} \mathcal{R} (\nu'/\nu_{c}').
\label{P_syn}
\end{equation}
By the same way, the total spectral energy distribution of SSC radiation is
\begin{equation}
P'_{\rm ssc}(\nu_{\rm ssc}')=\int \int \frac{dN_{\rm e}'}{d\gamma_{e}'} h \nu_{\rm ssc}'  \frac{dN_{\gamma}'}  {dt'd\nu_{\rm ssc}'} d\gamma_{e}'.
\label{P_ssc}
\end{equation}
The quantity $dN_{\gamma}'/dt'd\nu_{\rm ssc}'$ in Eq.(\ref{P_ssc}) is the scattered photon spectrum per electron\citep{blu70}, we can express $dN_{\gamma}'/dt'd\nu_{\rm ssc}'$ in terms of the frequency distribution of the seed photons $n_{\rm \nu'}$,
\begin{equation}
\frac{dN_{\gamma}'}{dt'd\nu_{\rm ssc}'}=\frac{3\sigma_{T}c}{4\gamma_{e}'^2} \frac{n_{\rm \nu'}d\nu'}{\nu'} F(q,g).
\label{spec_per}
\end{equation}

Given the synchrotron spectra $P'_{\rm syn} (\nu')$ and SSC spectra $P'_{\rm ssc} (\nu')$ in the co-moving frame, assumed the luminosity distance of the source from the observer is $D_L$, if we ignore the effect of the equal-arrival-time surface (EATS)\citep{wax1997, gra1999}, the intrinsic spectral flux at the observer frame can be expressed as
\begin{equation}
F_{\rm \nu, w.o.EATS} =\frac{(1+z) P'(\rm \nu'(\nu)) \Gamma }{4 \pi D_L^2},
\label{obs frame}
\end{equation}
where $\nu =\frac{4 \Gamma}{3(1+z)} \nu'$ is the observed frequency. But if we take the EATS effect into account, the respective intrinsic synchrotron and SSC fluxes at the observer frame are given by\citep{geng18}
\begin{equation}
F_{\rm \nu,  w.EATS} =\frac{1+z}{4 \pi D_{L}^2} \int_0^{\theta_j} P'(\nu'(\nu_{\rm obs})) D_{\rm dopp}^3 \frac{\sin \theta}{2} d\theta
\label{obs frame}
\end{equation}
where $\theta_j$ is the half-opening angle of the jet, $\nu_{\rm obs}=D_{\rm dopp} \nu'/(1+z)$ is the observed frequency when consider the EATS effect, and $D_{\rm dopp}=1/[\Gamma(1-\beta \cos \theta)]$ is the Doppler factor. 

In this work, the luminosity distance $D_L$ is obtained by adopting a flat $\rm \Lambda CDM$ universe, in which $H_0=71 \rm km s^{-1}$, $\Omega_m=0.27$, and $\Omega_{\rm \Lambda}=0.73$. 

\subsection{Absorption}

The high-energy gamma-rays may suffer the internal absorption by ambient photons inside the source and external absorption by extragalactic background light (EBL) during they travel through the interstellar medium. When we consider the internal absorption by the source, the high energy gamma-rays can be attenuated due to interaction with synchrotron photons of the afterglow phase through the pair production effect. The optical depth for a gamma-ray with energy $\varepsilon_\gamma=\frac{4\Gamma}{3(1+z)}\varepsilon_\gamma'$ is expressed as \citep{ver2016, murase2011}

\begin{equation}
\tau_{\rm in} = \frac{\Delta'}{2} \int_{-1}^{-1} d\mu (1-\mu) \int d \nu' n_{\nu'} \sigma_{\rm \gamma\gamma},
\label{tau_internal}
\end{equation}
where 

\begin{equation}
\sigma_{\gamma \gamma} = \frac{3}{16} \sigma_T (1-\beta_{\rm cm}^2) \times \nonumber\\
 \left[2 \beta_{\rm cm} (\beta_{\rm cm}^2-2) + (3 - \beta_{\rm cm}^4) \ln \frac{1+\beta_{\rm cm}}{1-\beta_{\rm cm}}\right],
\label{sigma}
\end{equation}
is the pair production cross section,  $\beta_{\rm cm}$ is the velocity of the $e^{\pm}$ in the center of mass frame, i.e.,
\begin{equation}
\beta_{\rm cm}=\frac{v}{c}=\sqrt{1-\frac{2(m_e c^2)^2}{\varepsilon'_{\gamma} \varepsilon'_{t} (1-\mu)}},
\end{equation}
$\mu=\cos \theta_i$, $\theta_i$ is the angle between the directions of the interacting photons, $\varepsilon'_{t}=h\nu'$ is the energy of a target photon (synchrotron photon). $\Delta'$ can't be cancelled out in Eq.(\ref{tau_internal}) and we take $\Delta'\simeq R/10\Gamma$, we assume that the emitting electrons are homogeneously distributed in the shell.

If the intrinsic flux of a source is $F_{\nu}$ and the flux after the internal $\gamma \gamma$ absorption is $F_{\nu}^{\gamma \gamma}$, then we have
\begin{equation}
F_{\nu}^{\gamma \gamma}=F_{\nu} \left[(1-\exp(-\tau_{\rm in}))/\tau_{\rm in}\right].
\end{equation} 

When we consider the absorption by EBL, the optical depth $\tau_{\rm EBL}$ versus observed energy of gamma-ray photons at different redshifts $z$ are given by \citet{dom2011}, after absorpting by EBL, the observed flux $F_{\nu}^{\rm obs}$ is
\begin{equation}
F_{\nu}^{\rm obs}=F_{\nu}^{\gamma \gamma} \exp(-\tau_{\rm EBL}).
\end{equation}

\section{Numerical appoach}

In order to calculate the time-dependent spectra of high-energy afterglow emission, it is necessary to solve the hydrodynamic evolution of GRB shock. The parameters set as $E_k=3 \times 10^{54} \rm erg$, $\Gamma_0=200$ and $n=0.1 \rm cm^{-3}$. Numerical solving the Eq. (\ref{dyn1}) - Eq. (\ref{dyn3}), we obtain three dynamic phases of the shock: (1) coasting phase: at $t \lesssim t_{\rm dec}$, the shock propagates at a constant velocity $\Gamma_0$ and the shock radius is $R \propto t$; (2) Blandford-McKee self-similar phase: at $t \gtrsim t_{\rm dec}$, the shock dynamics transits into the self-similar Blandford-McKee solution, the shock Lorentz factor is $\Gamma \propto t^{-3/8}$ and the shock radius is $R \propto t^{1/4}$\citep{bla76, sari1998}. The deceleration timescale is close to the peak time of the r-band light curve, i.e., $t_{\rm dec} \sim 10^2 \rm s$; (3) Sedov-Taylor self-similar phase: at $t \gg t_{\rm dec}$, the shock dynamics transits into the non-relativistic outflow, then we can approximate that $\beta \propto t^{-3/5}$ and $R \propto t^{2/5}$\citep{sedov1959, taylor1950}. It should be emphasized that the hydrodynamic evolution in this work is available for strong shocks not only the relativistic, but also the non-relativistic, or mildly relativistic regime.

In general, the display difference scheme (DDS) is usually used to solve the continuity equation of electrons, but when we discretize the time and energy space, the time step and the energy step have to follow the Courant-Friedrichs-Lewy (CFL) condition, otherwise we can't ensure the convergance of the solution for the arbitrary values of time steps. To avoid this problem, we must adopt an unconditional stable scheme: the fully implicit difference scheme (FIDS), which is proposed by \citet{chang1970, chia1999}. The FIDS can find a more stable, non-negative and particle number-conserving solution, reduce the number of mesh points and obtain accurate solutions. For more details on this numerical methods, please see \citep{chia1999}. Here, we simply present the discretization procedures to solve the Eq.(\ref{continuity equation}). The Lorentz factor $\gamma_e'$ of the ED ranges from $10$ to $10^9$, it is necessary to use an energy grid with equal logarithmic resolution. Then the oberved time step for every evolution ranges from $10 \rm s$ to $10^7 \rm s$, also with equal logarithmic resolution. After obtaining the solved-out temporal evolution of the ED $dN_{e}' / d\gamma_{e}'$, the total synchrotron and SSC radiation power in the comoving frame are calculated by Eq.(\ref{P_syn}) and Eq.(\ref{P_ssc}). In the performed simulations a grid of 500 points has been used both for electron Lorentz factor and photon frequency, and a grid of 121 points has been used for observed time.

\begin{figure}
\vskip -0.0 true cm
\centering
\includegraphics[scale=0.45]{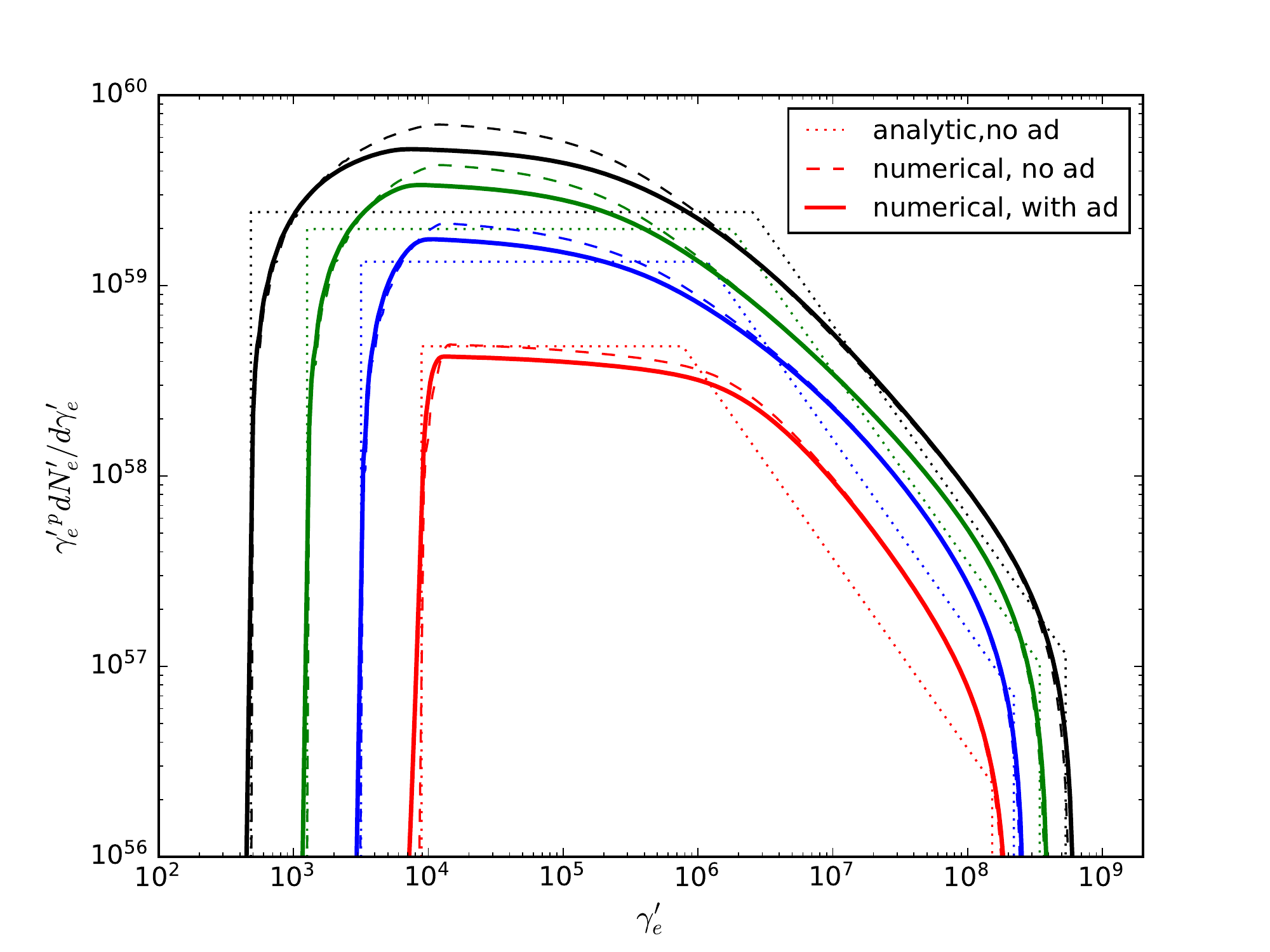}
\includegraphics[scale=0.45]{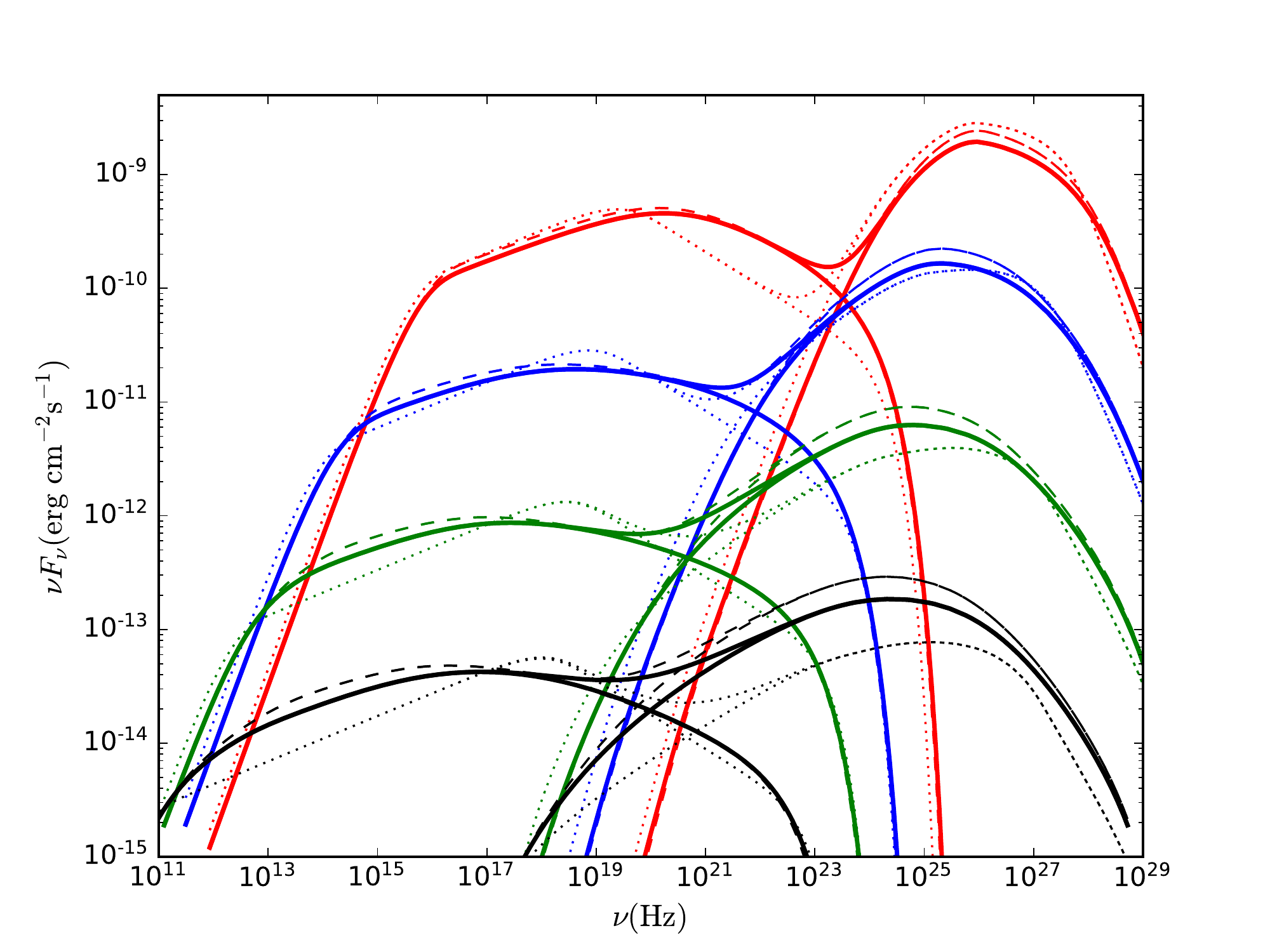}
\caption{The EDs in the fluid frame (upper panel) and the radiation spectra in the observer frame (lower panel) at different times ($t=10^2 \rm s$ (red lines), $t=10^3 \rm s$ (blue lines), $t=10^4 \rm s$ (green lines), $t=10^5 \rm s$ (black lines)). The simulation parameters are $E_k=1 \times 10^{54} \rm erg$, $\Gamma_0=200$, $n=1 \rm cm^{-3}$, $\epsilon_e=0.1$, $\epsilon_B=1 \times 10^{-5}$ and $p=2.6$, the GRB is located at a redshift $z=1$.}
\label{Fig:Compare}
\end{figure}

In order to test the validity of our code, we have compared our simulation results with those obtained by \citet{wang2010}, who have developed a similar analytical method, but adiabatic cooling was not considered. The EDs (in the comoving frame) and the radiation spectra (in the observed frame, without considering the EATS effect and absorption) at different observed times ($t=10^2 \rm s$, $t=10^3 \rm s$, $t=10^4 \rm s$, $t=10^5 \rm s$) are shown in Fig.\ref{Fig:Compare}. The results are obtained with the same set of parameters. We can see that the two methods produce EDs and radiation spectra with very similar shapes. The high energy electrons are cooled above $\gamma_c'$ ($\gamma_c'$ is the cooling Lorentz factor), and the EDs are going as $dN_{e}' / d\gamma_{e}' \propto \gamma_e'^{-p-1}$. The low energy electrons are uncooled between $\gamma_m'$ and $\gamma_c'$, and the EDs are gradually approaching the slope of the injected function, i.e., $dN_{e}' / d\gamma_{e}' \propto \gamma_e'^{-p}$. The results indicate that the EDs should be expressed with the slow cooling approximation for $t \gtrsim 10^2 \rm s$. The analytical solution obtains a pure broken power-law segment with sharp break, that deviates from the numerical model. With time evolution, we can see that the normalization of numerical method is slightly different from \citet{wang2010}, the corresponding cooling Lorentz factor $\gamma_c'$ is also slightly different, but the deviations are within half an order of magnitude. The simple broken-power-law function is not appropriate for the numerical results. We suspect this differences arise from the fact that our numerical model considers a time-dependent electron distribution, but the analytical model considers the electron distribution at a certain moment and ignores the details of the intermediate evolution. To study whether the adiabatic cooling effects the electron distribution, we turned off this process for test simulation, but there has no visible impact on the electron distribution and radiation spectra. Based on the consistency of the results, we can conclude that the FIDS is a good method to solve the continuity equation to ensure the convergence of the solution in the intermediate steps.

\section{Apply to GRB 201216C}

\begin{deluxetable*}{cccccccc}
\tabletypesize{\scriptsize}
%\rotate
\tablecolumns{8}
\tablewidth{0pc}
\tablecaption{Model parameters for the afterglow emission of GRB 201216C.}
\tablehead{ \colhead{Model} &  \colhead{$E_k$ [$\rm erg$]} & \colhead{$\Gamma_0$} & \colhead{$\epsilon_{e}$}  & \colhead{$\epsilon_{B}$}  & \colhead{$n$ [$\rm cm^{-3}$]} & \colhead{$p$} & \colhead{EATS} }
\startdata
A & $3 \times 10^{54}$ & 200 & 0.7 & $6 \times 10^{-6}$ & 0.1 & 2.6 & no \\
B & $6 \times 10^{55}$ & 100 & 0.02 & $2 \times 10^{-5}$ & 1 & 2.6 & no \\
C & $6 \times 10^{53}$ & 100 & 0.6 & $8 \times 10^{-5}$ & 0.5 & 2.1 & yes, $\theta_j=0.05$\\
\enddata
\label{Tab:tab1}
\end{deluxetable*}

In this section, we perform modelings of the availabel multi-wavelength data for GRB 201216C. The modelings are based on \S 2 and the numerical methods are based on \S 3. Since that there are only a few measurements and the model has 6 (without considering the EATS effect) or 7 (with considering the EATS effect) parameters plus very detailed physics. We consider several sets of parameters from the following reasons:
\begin{itemize}
\item The Fermi-GBM observations implies that the isotropic energy release of this GRB is $E_{\rm iso} \sim 5 \times 10^{53} \rm erg$. Assume that the radiative efficiency for producing prompt gamma-ray emission is $1 \%-100 \%$, then we can obtain the kinetic energy of the outflow $E_{\rm k} \sim 5 \times 10^{53} \rm erg- 5 \times 10^{55} \rm erg$.
\item Assume an ultra-relativistic outflow, we take the initial Lorentz factor $\Gamma_0 \sim 100$.
\item $n$ is the external medium density of GRBs, for long GRB scenario, $n$ is taken as $\sim 1 \rm cm^{-3}$.
\item The observed peak time of r-band light curve is $t_p \sim 177 \rm s$. Assume that $\Gamma_0 \sim 100$ and the peak time equals the deceleration timescale of the shock, then we can obtain a rough relationship between $n$ and $E_k$. With the smaller $n$ is, the larger $E_k$ is.
\item The observed r-band light curve shows a power-law decay, i.e., $F_r \propto t^{-1.1}$. Assume a ISM scenario, the r-band (low frequency) light curve evolution for a standard GRB afterglow is $F_{\nu}(t) \propto t^{3(1-p)/4}$ (late time) or $F_{\nu}(t) \propto t^{(2-3p)/4}$ (very late time) \citep{sari1998}, then we obtain the corresponding electron distribution index $p \sim 2.5$ or $p \sim 2.1$.
\item The values of $\epsilon_e$ and $\epsilon_B$ are set by the micro-physics of relativistic shocks. From the literature, $\epsilon_e$ ranges from $0.02-0.6$ and $\epsilon_B$ ranges from $10^{-6} -10^{-3}$\citep{santana2014}.
\item Since no jet breaks are observed in X-ray and r-band light curves until $t\sim 10^6 \rm s$, the afterglow radiation model described so far is based on the assumption of an on-axis viewing angle and the ejecta is almost isotropic. Therefore, we don't consider the EATS effect in model A and model B.
\item However, it is also possible that the jet breaks are not obvious or not be observed. In this case, we should consider the EATS effect in our calculation, and add one more parameter $\theta_j$ in model C.
\end{itemize}

\begin{figure}
\vskip -0.0 true cm
\centering
\includegraphics[scale=0.45]{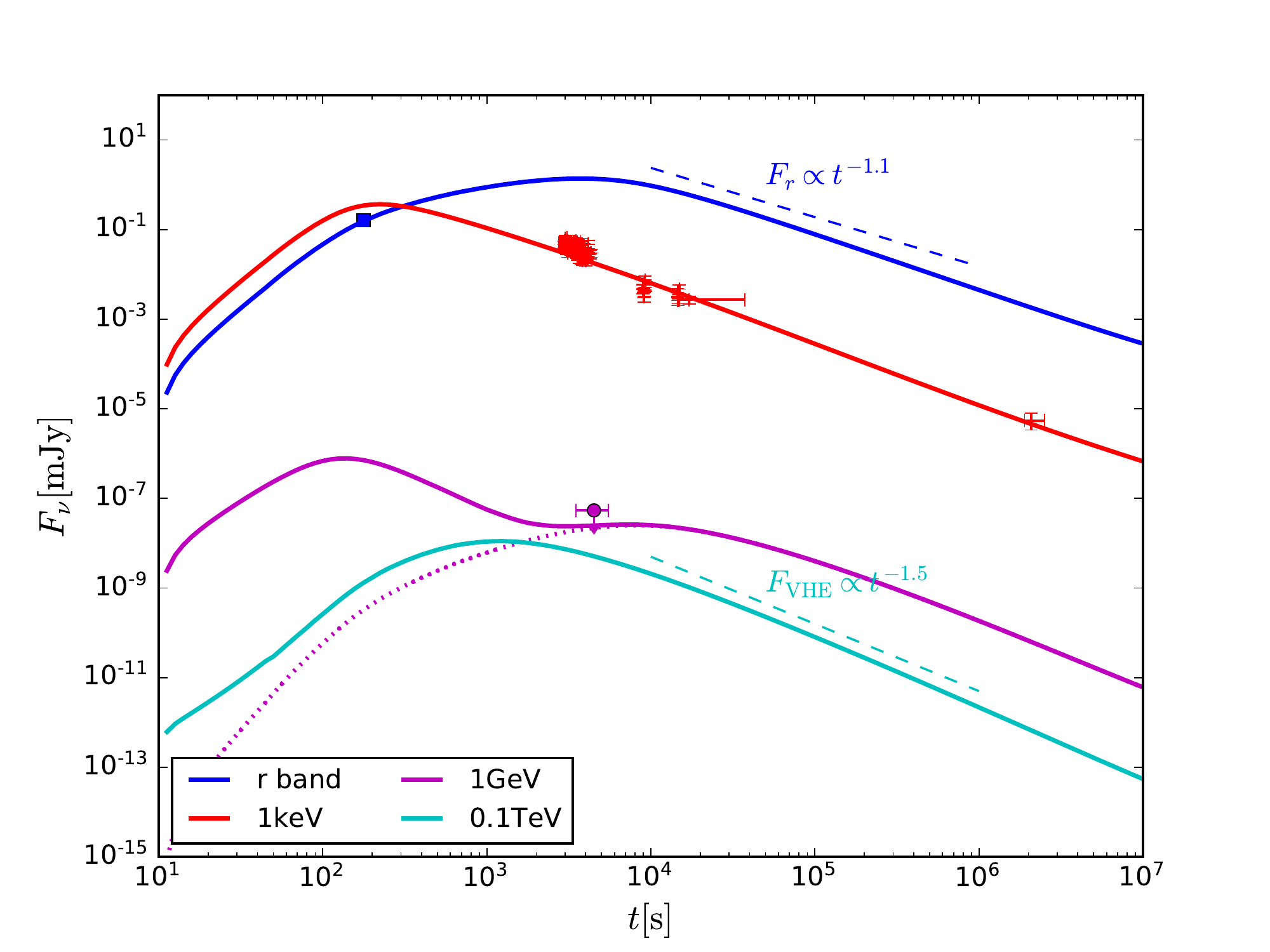}
\includegraphics[scale=0.45]{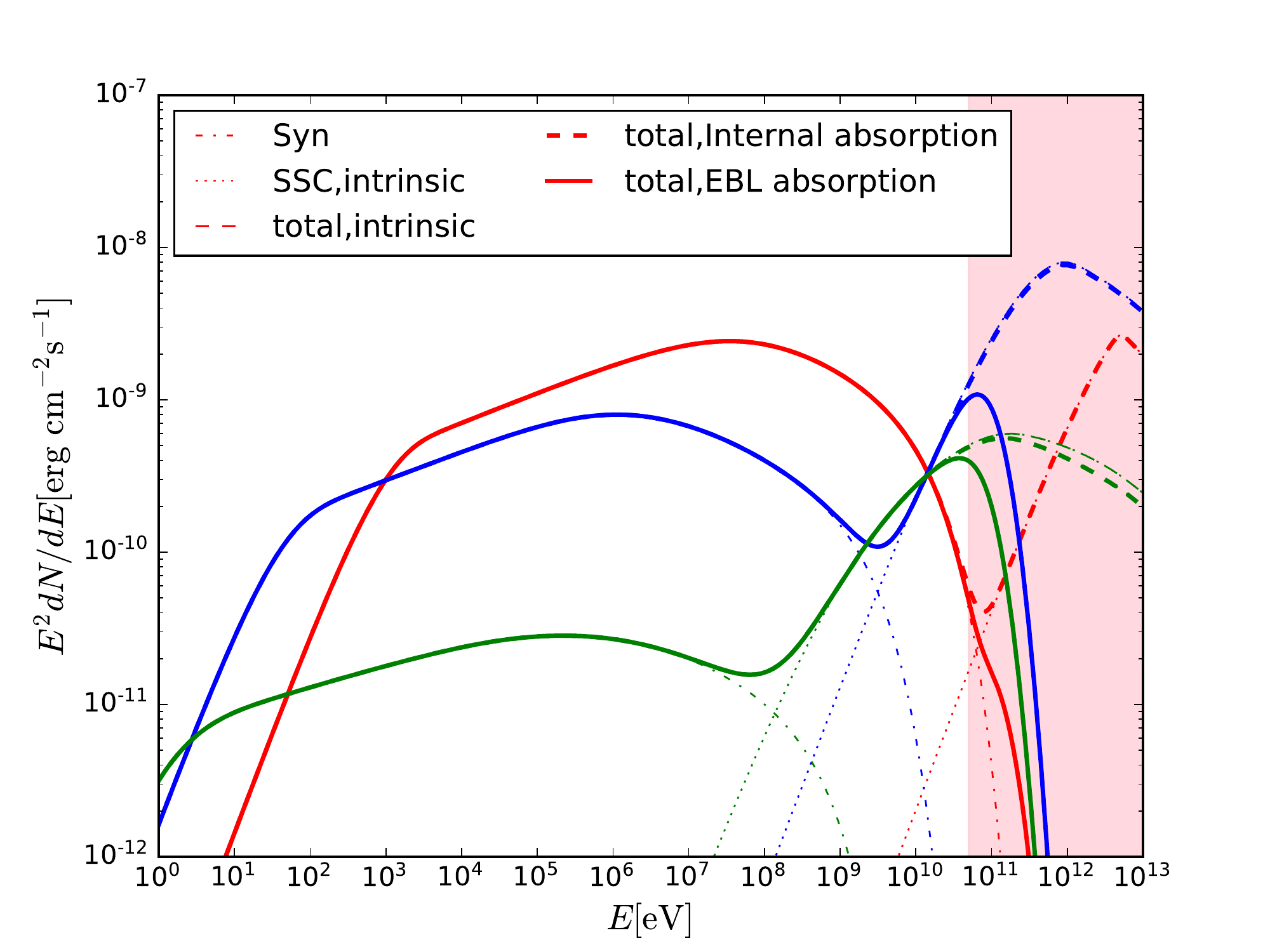}
\caption{Light curves and spectra of model A for GRB 201216C. Upper panel: The theoretical intrinsic light curves at optical (r-band), X-ray (1 keV), gamma-ray (1 GeV) to Sub-TeV (0.1 TeV) band, then compare with the observations. The optical data is taken from \citet{shre2020}, the X-ray data are taken from \citet{evans2021} by assuming a power-law spectrum with photon index $\Gamma_{X}=2.35$ and the Fermi-LAT (1 GeV) data are taken from \citet{biss2020} by assuming a power-law spectral with photon index $\Gamma_{\rm LAT}=2$. The purple dotted line show 1GeV light curve which is obtained by only switching on the SSC emission artificially. 
Lower panel: spectra of synchrotron, SSC and total emission with or without absorption at $t \sim 10^2 \rm s$ (red lines), $t \sim 10^3 \rm s$ (blue lines) and $t \sim 10^4 \rm s$ (green lines). The pink shaded region is the energy range of the MAGIC telescope (50 GeV to 10 TeV).}
\label{Fig:GRB_LC1}
\end{figure}

\begin{figure}
\vskip -0.0 true cm
\centering
\includegraphics[scale=0.45]{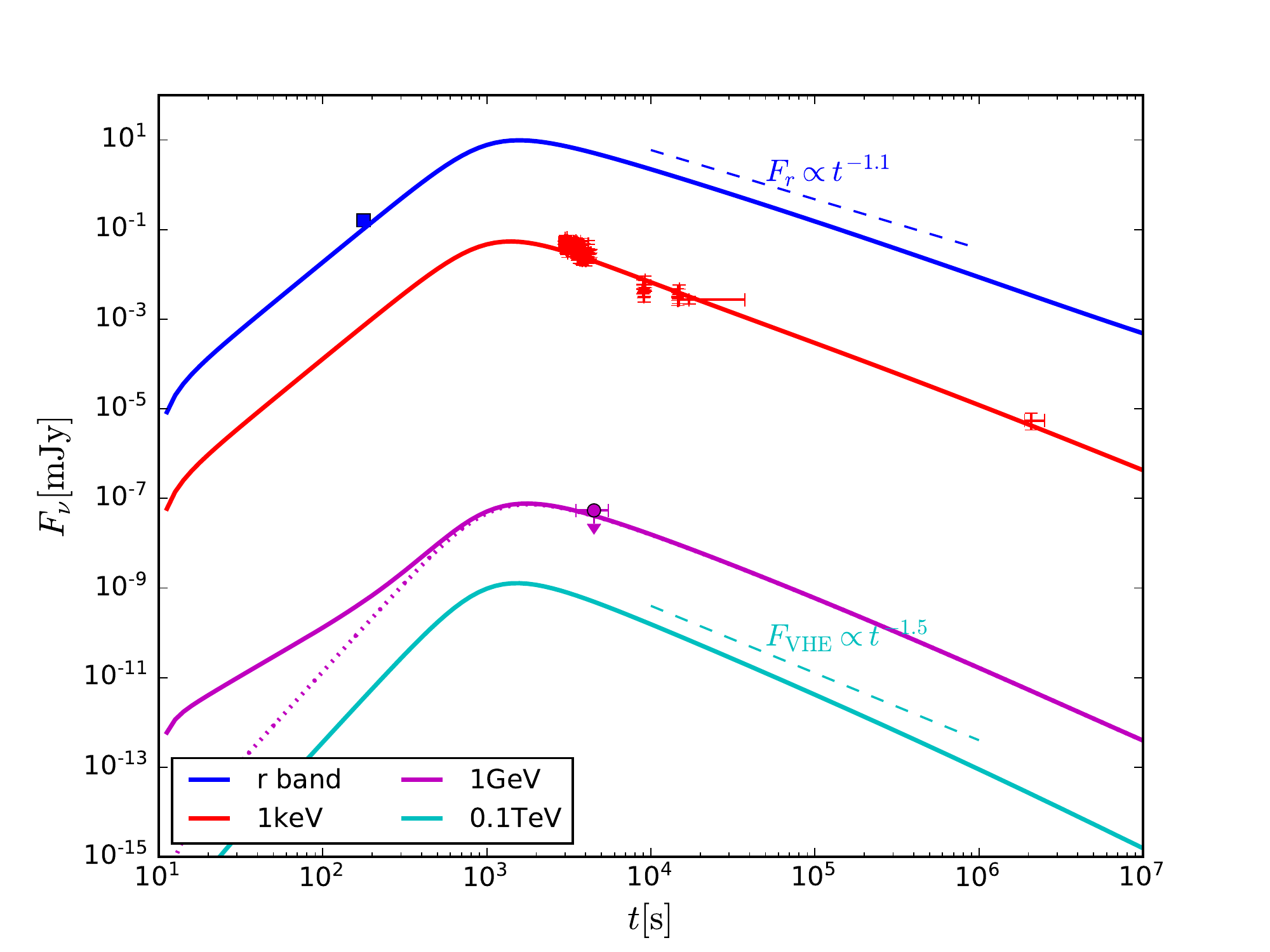}
\includegraphics[scale=0.45]{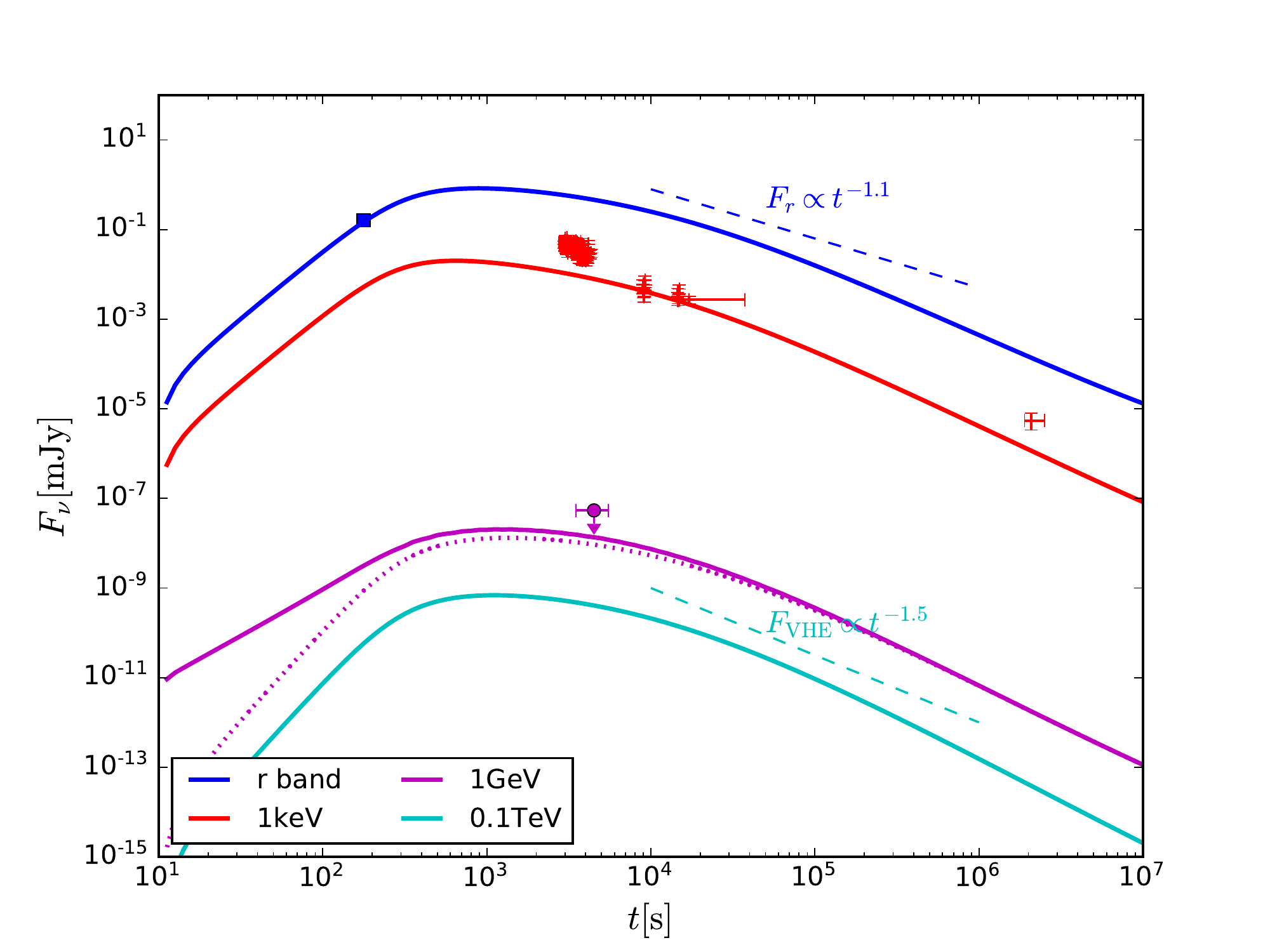}
\caption{Light curves of model B (upper panel) and model C (lower panel) for GRB 201216C. The data points are the same as in Figure \ref{Fig:GRB_LC1} and the model light curves are plotted with solid lines. The purple dotted line are 1GeV light curve which is switching on the SSC emission artificially. }
\label{Fig:GRB_LC2}
\end{figure}

On the basis of above discussions, we manually adjust the 6 (or 7) parameters of our numerical model to explain the observed multi-wavelength light curves data by naked eyes. We have tested three models, whose parameters are summarized in Table.\ref{Tab:tab1} and the modeling results are shown in Fig.\ref{Fig:GRB_LC1} and Fig.\ref{Fig:GRB_LC2}.

In model A, we consider the large $\epsilon_e$ case and the results are shown in Fig.\ref{Fig:GRB_LC1}. In the upper panel of Fig.\ref{Fig:GRB_LC1}, the solid lines show the theoretical multi-wavelength light curves, then compare with the observations. We can see that the observed X-ray and r-band light curves data are well explained by model A. The predicted light curve of r-band shows a power law decay in flux with its decay index is about $-1.1$ and the peak time of r-band light curve is about a few hundred seconds, consist with the observations of LT, VLT and FRAM-ORM\citep{shre2020, jeli2020}. The predicted flux of 1 GeV at $10^3 -10^4 ~\rm s$ is $F_{\nu} \sim 10^{-8} \rm mJy$, then $\nu F_{\nu} \sim 10^{-11} \rm erg~cm^{-2}~s^{-1}$, it is too low to be detected by Fermi-LAT. But if Fermi-LAT is sensitive enough and this GRB was on the Fermi-LAT field of view, we will see the emission mechanism is switching from synchrotron to SSC at $t \sim 10^3 \rm s$, i.e., the early light curve of 1 GeV is contributed by the synchrotron radiation, while the late time light curve of 1 GeV is contributed by the SSC radiation. The light curve of 1 GeV can't be a smooth power law. We are more concerned about the VHE emissions, but the detail observations of MAGIC telescope have not been published yet. We only known that the VHE emissions of this burst has a $\sim 5 \sigma$ statistical significance excess. If we don't consider any absorption and just consider the intrinsic flux of this GRB, the predicted light curve of 0.1 TeV is the cyan solid line. We can see that the light curve of 0.1 TeV has a peak flux $F_{\nu} \sim 10^{-8} \rm mJy$, then $\nu F_{\nu} \sim 10^{-9} \rm erg~cm^{-2}~s^{-1}$, and the peak time of 0.1 TeV light curve is $\sim 10^3 \rm s$. After the peak flux, the light curve decays as the power law with $F_{\rm VHE} \propto t^{-1.5}$. 
%In the early stage ($t \lesssim \rm 100s$), compare with other band light curve, the light curve behavior of 0.1 TeV shows a slower rise, then follows a normal rise. It is suggested that the radiation mechanism of 0.1 TeV may be switching at $t \lesssim \rm 100s$. 

In the lower panel of Fig.\ref{Fig:GRB_LC1}, we show the spectral energy distributions (SEDs) of synchrotron, SSC and total emissions with or without absorption at $t \sim 10^{2} \rm s$, $t \sim 10^{3} \rm s$ and $t \sim 10^{4} \rm s$ for model A. The synchrotron component contributes dominantly to GeV emissions at $t \sim 10^2 \rm s$ and the flux of 1 GeV is $\gtrsim 10^{-9} \rm erg~cm^{-2}~s^{-1}$. Unfortunately, this GRB was not in the field of view of Fermi-LAT at $t \sim 10^2 \rm s$. The maximum energy of synchrotron radiation can reach $\sim$10 GeV at $t \sim 10^2 \rm s$ and the flux decays exponentially with energy above 10 GeV. The SED at $t=10^3 \rm s$ shows an obvious transition from synchrotron component to SSC component around $\sim$ GeV at $t=10^3 \rm s$. Then the SSC component contributes dominantly to the GeV emissions at $t=10^4 \rm s$ and the flux keep rising with the energy above GeV. However, the flux of GeV is only $\sim 10^{-10} \rm erg~cm^{-2}~s^{-1}$ at $t \sim 10^3 -10^4 \rm s$ and the current sensitivity of Fermi-LAT is not good enough to detect this evolution of SEDs. What's exciting is that the expected Sub-TeV flux of this GRB is comparable to the GeV flux, which can be explained the detection of MAGIC. For the VHE energy range, we can see that the internal absorption can hardly changes the energy spectra of SSC component, but the EBL absorbs most of SSC emissions, especially above 0.3 TeV. However, the emissions below 0.1 TeV are not completely absorbed by EBL. Therefore, as long as the emissions below 0.1 TeV are strong enough, the VHE telescope can still observe the low energy VHE signal.

The small value of $\epsilon_e$ leads to a large value of $E_k$. In model B, we consider the small value of $\epsilon_e$, and the fitting results are shown in the upper panel of Fig.\ref{Fig:GRB_LC2}. We can see that Model B can explain the observed X-ray and optical data points, as well as the decay slope of r-band and 0.1 TeV light curves. Howerver, the predicted flux of 0.1 GeV light curve is higher and has already reached the upper limit values of Fermi-LAT observation at $10^3 \rm s$.
%If Fermi-LAT is observing at this time, this source should be observable.

If we consider the EATS effect, the fitting results are shown in the lower panel of Fig.\ref{Fig:GRB_LC2}. We can see that Model C can explain the observed optical data point, the Fermi-LAT upper limit value and the decay slope of 0.1 TeV light curve. However, the predicted X-ray light curve is slightly lower than the observed data, and the power-law decay index of r-band light curve is steeper than $-1.1$. In addiation, we obtain a relative small jet half-opening angle, i.e., $\theta_j=0.05$ rad, indicate that the kinetic energy of the shock is smaller than model A and model B.

\section{Discussions and conclusions}

In this paper, we performed a self-consistent numerical modeling to calculate the temporal evolutions of the energy distributions of electrons and photons in the GRB shock, then we apply our numerical model to explain the available multi-wavelength observations of distant GRB 201216C. The observed optical and X-ray afterglow emissions can be explained by the synchrotron radiation, while the Sub-TeV emissions are dominated by the SSC radiation. The physical parameters of this GRB are still uncertain due to the limited observations and the parameter values used in Table. \ref{Tab:tab1} are optimized for explaining the optical and X-ray light curve, and the Fermi-LAT upper limit. But anyway, we offer some possibility to explain the available observations for this GRB. 

The afterglow model described so far is based on the assumption of ISM model. We found that all the models in Table. \ref{Tab:tab1} have a very small value of $\epsilon_B$, the low value of $\epsilon_B$ leads to a slow cooling regime for the shocked electrons, i.e., $\gamma_m'<\gamma_c'$, and the observed optical frequency is away between $\nu_m$ and $\nu_c$ ($\nu_m$ and $\nu_c$ are the characteristic frequencies of $\gamma_m'$ and $\gamma_c'$). If we consider the wind-like scenario, the late time afterglow light curve evolution of r-band is $F_{\nu}(t) \propto t^{(1-3p)/4}$\citep{chevalier2000}. Assume $p \sim (-3 \to -2)$, then the decay rate of the r-band light curve is $\alpha_{\rm r} \sim (-2 \to -1.25)$, more steeper than the observed decay rate $\alpha_{\rm r} \sim -1.1$. But the ISM model can explain this decay slope (show in Fig.\ref{Fig:GRB_LC1} and Fig.\ref{Fig:GRB_LC2}). Also, our model predicts the density of the external medium is relatively low, such low medium density does not consistent with the wind-like scenario.

\begin{figure}
\vskip -0.0 true cm
\centering
\includegraphics[scale=0.6]{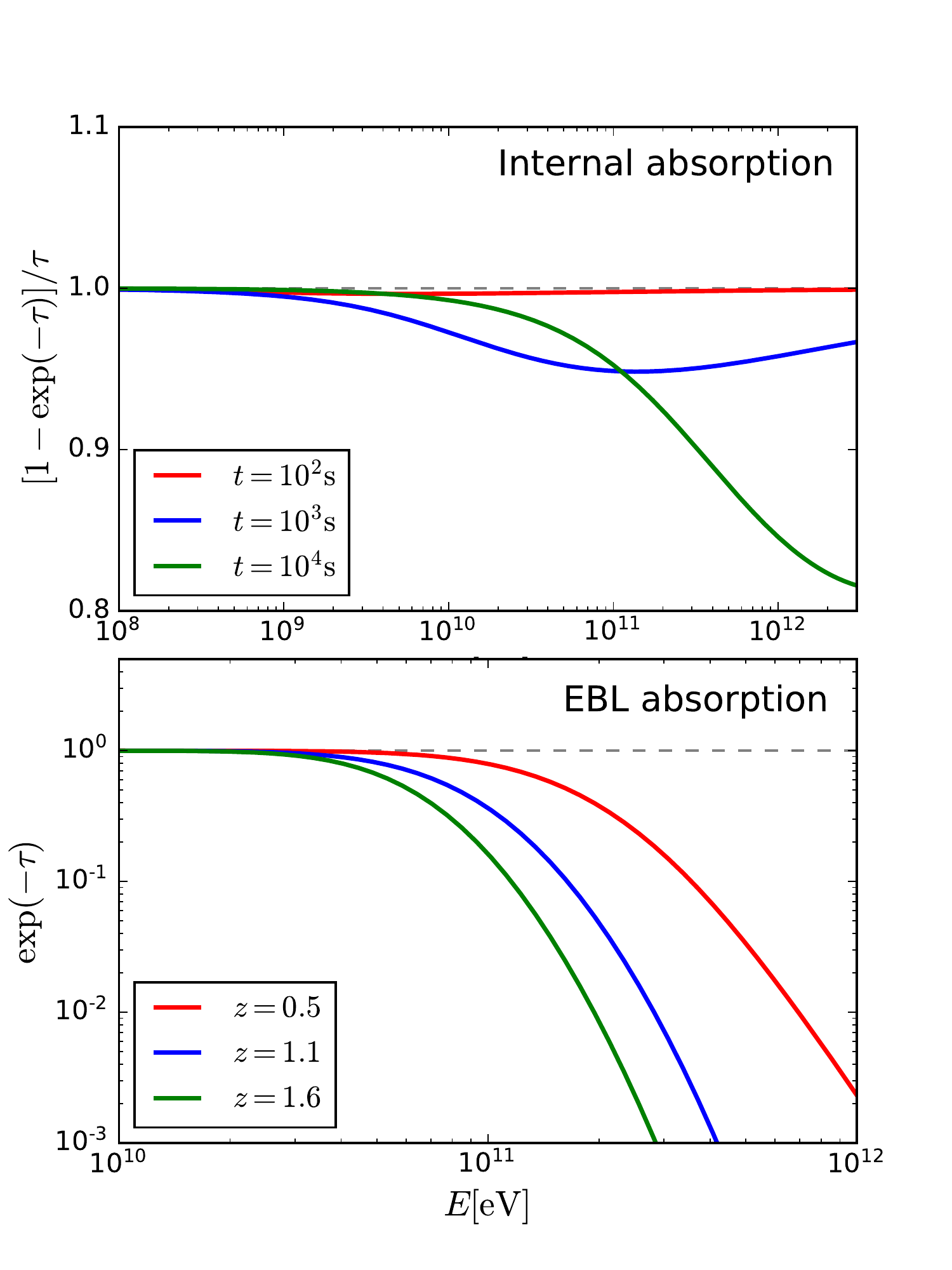}
\caption{Upper panel: The internal absorption ($[1-\exp(-\tau_{\rm in})]/\tau_{\rm in}$) at different times, with its relevant physical parameters are the same as Fig.\ref{Fig:GRB_LC1}. 
Lower panel: The EBL absorption ($\exp(-\tau_{\rm EBL})$) at different redshifts.}
\label{Fig:EBL}
\end{figure}

\begin{figure}
\vskip -0.0 true cm
\centering
\includegraphics[scale=0.6]{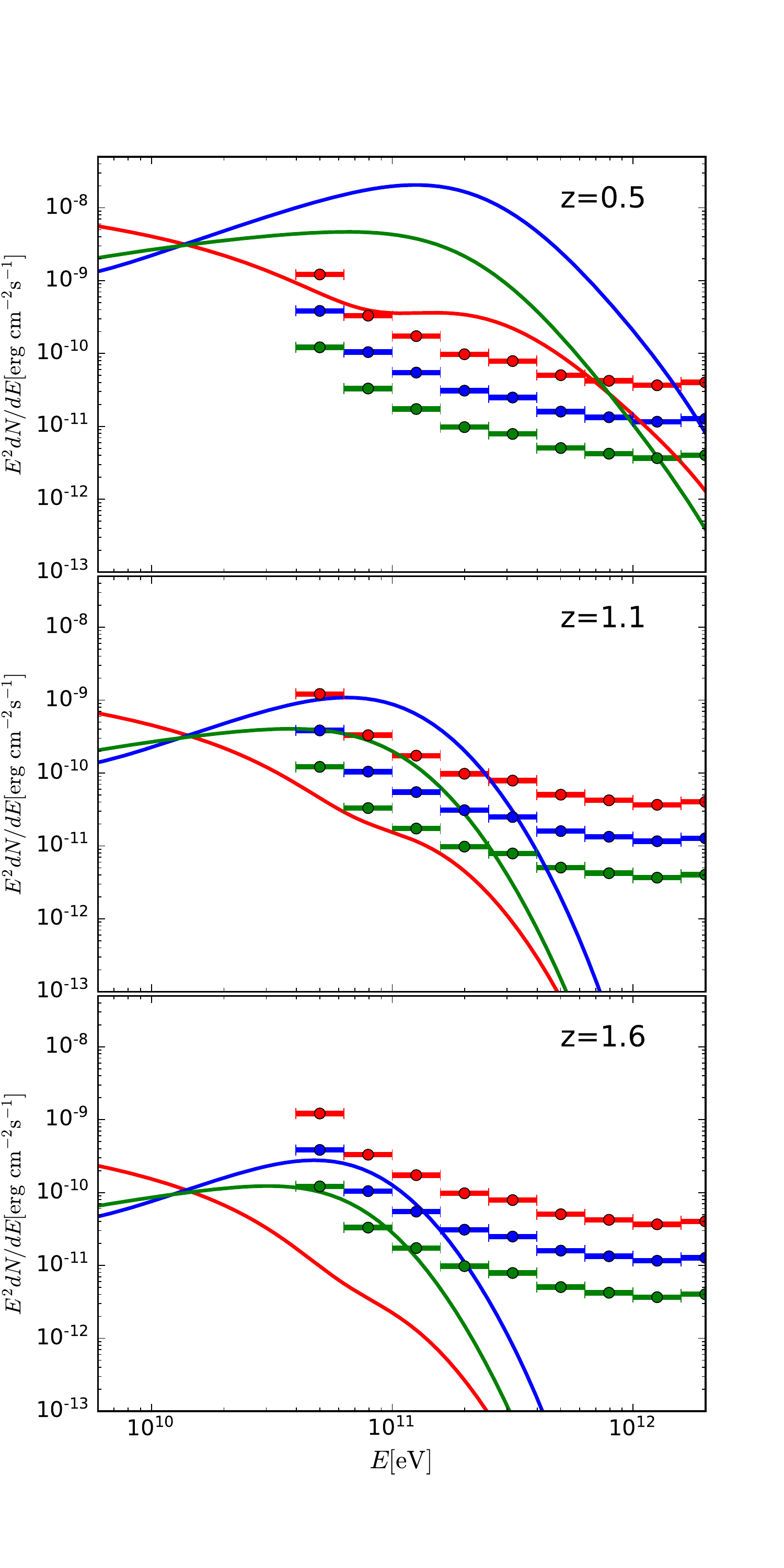}
\caption{We take the parameters of model A as the typical values and compare the observed VHE spectral distribution at different times (red solid lines for $t \sim 10^2 \rm s$, blue sold lines for $t \sim 10^3 \rm s$ and green sold lines for $t \sim 10^4 \rm s$) and at different redshifts (upper panel for $z=0.5$, middle panel for $z=1.1$ and lower panel for $z=1.6$). The differential sensitivity of MAGIC scaled to duration time of observation, i.e., $10^2\rm s$ (red data points), $10^3 \rm s$ (blue data points) and $10^4 \rm s$ (green data points) are also shown. The differential sensitivity of MAGIC are adopted from \url{https://magic.mpp.mpg.de/uploads/pics/info.txt}.}
\label{Fig: sensitivity}
\end{figure}

To be noticed that when we dealing with the continuity equation of electrons, the creation and annihilation of pairs wasn't considered in our numerical calculation. This is because the cooling effects of the pairs creation and annihilation have no observable impact on the afterglow emissions. We can obtain this conclusion simply from the internal absorption of VHE emissions. The effects of EBL absorption together with that of internal absorption derived from model A are visualised in Fig.\ref{Fig:EBL}. The upper panel of Fig.\ref{Fig:EBL} shows the internal absorption at different times. The results suggest that the internal absorption is minimal in the VHE energy range. Even at $t=10^4 \rm s$, the internal absorption only absorbs $\lesssim 20\%$ emissions at 1TeV. The internal absorption is even less at lower energy and earlier time. The lower panel of Fig.\ref{Fig:EBL} shows the EBL absorption at different redshifts. The results suggest that with the increasing of redshift, the EBL absorption becomes stronger. The EBL absorption occupies a major effect for energies up to 30 GeV. For energies up to 0.1 TeV, the observed flux after EBL absorption are 0.8, 0.4 and 0.2 times of the original flux for $z=$ 0.5, 1.1 and 1.6 respectively. For energies up to 0.3 TeV, the observed flux after EBL absorption are only $0.2$, $10^{-2}$ and $10^{-3}$ times of the original flux for $z=$ 0.5, 1.1 and 1.6, it implies the emissions of distant GRBs ($z \gtrsim 1$) are almost completely absorbed by EBL. Hence, for distant GRBs, we don't expect to observe the emissions above 0.3 TeV.

We can estimate the maximum distance of GRBs that can be detected by MAGIC. The specific approaches are as follows: Firstly, assume that GRB 201216C is a typical VHE GRB and the parameters of model A are the typical intrinsic parameters; Secondly, place GRB 201216C at different cosmological distances artificially and calculate the corresponding observed flux; Thirdly, compare the observed flux with the differential sensitivity\footnote{In the medium range of duration times, the sensitivity of MAGIC telescope follows the usual $\propto 1/\sqrt{t}$ dependence\citep{ale2016}.} of MAGIC for different duration time. If the observed flux is higher than the sensitivity, the VHE emissions can be observed by MAGIC; otherwise, this GRB can’t be observed. The results are shown in Fig.\ref{Fig: sensitivity}. For the case of low redshift ($z \sim 0.5$), the GRB has a higher flux and less EBL absorption. The spectral flux observed at different time is higher than the corresponding sensitivity, so the early and late time ($\sim 10^2-10^4 \rm s$) VHE emissions can be observed by MAGIC, and the maximum energy of VHE photons can reach the order of several TeV. In the case of medium redshift (GRB 201216C, or $z \sim 1.1$), due to good sensitivity of MAGIC, it is likely that the $\lesssim 0.3 \rm ~ TeV$ emissions of GRB can be observed, especially for the observation at $10^3-10^4 \rm s$. But the early ($t \lesssim 100 \rm s$) Sub-TeV emissions may be invisible by MAGIC. The maximum energy of VHE photons can only reach a few hundred GeV. In the case of high redshift ($z \sim 1.6$), the GRB flux is weaker and EBL absorption is even stronger. We can see that the observed VHE flux exactly equals to the sensitivity of MAGIC. We believe that if the distance of this GRB is farther, it can’t be observed by MAGIC. Therefore, we concluded that the GRB falls below MAGIC sensitivity at $z \sim 1.6$ ($D_L \sim 10 \rm Gpc$). The local event rate of long GRBs is $\rho_0 \sim 1 \rm Gpc^{-3}~yr^{-1}$\citep{wander2010}, then we can estimate that $\sim 1000$ GRBs are occured every year for $z<1.6$. Howerver, only a few  of VHE GRBs have been detected at present, suggesting that the generation and observation of VHE GRBs need extremely strict conditions.

Compare with two other normal GRBs which Sub-TeV emissions were also detected, such as GRB 190114C, GRB 180720B\footnote{GRB 190829A was also observed the VHE emissions by H.E.S.S, but several indications implied that GRB 190829A is a low luminosity GRBs\citep{hess2021, zhang2020, chand2020, lulu2021}, different from GRB 201216C.}, although GRB 201216C has a high redshift and we can't avoid the EBL absorption, but they share some common features\citep{huang2020} (1) Both GRBs are energetic enough, the kinetic energy of these three GRBs can reach $E_k=10^{53} - 10^{54} \rm erg$, it is 1-2 orders of magnitude higher than most normal long GRBs; (2) The external medium density of both shock is relative low, $n \sim 0.1-1 \rm cm^{-3}$. It is likely a uniform density ISM, rather than a stratified stellar wind, and such low medium density can reduce the internal absorption of the source and make the Sub-TeV emissions easier to be detected; (3) The outflow of both GRBs are ultra-relativistic with its initial bulk Lorentz factor are $100-600$, and both GRBs has a relativistic electron power law index with $p=2.2-2.6$; (4) The modeling gives a small magnetic equipartition factor of $\epsilon_B \sim 10^{-4}-10^{-6}$ and a relative large value of the electron energy density fraction of $\epsilon_e=0.1-0.7$. The case of $\epsilon_B \ll \epsilon_e$ suggests that a significant SSC component is expected\citep{zhang2001}. Such small value of $\epsilon_B$ also implies that the magnetic field may decay with distance downstream of the shock front\citep{lemoine2013a, lemoine2013b, lemoine2015}.

High redshift is an unique feature for GRB 201216C. Even with strong EBL absorption, its Sub-TeV emissions can still be observed. The observable VHE GRBs may be required the following special conditions: (1) The internal absorption can be avoided. The EBL absorption is not very important at $\sim$ 50 GeV, if the internal absorption can be avoided, then it is possible to observe the VHE emissions by MAGIC around 50 GeV-100 GeV. (2) The GRBs are energetic enough and the SSC component are significant enough, even with strong EBL absorption, such Sub-TeV emissions can still be observed. (3) The sensitivity of VHE telescope is good enough at lower energy, improve the low energy sensitivity of the VHE telescope is very important to detect the Sub-TeV emissions of these high redshift GRBs.

\acknowledgements{We thank Zhuo Li and Tianqi Huang for helpful comments and discussions. This work is supported in part by the National Key R$\&$D Program of China (2021YFC2203100), Anhui project (Z010118169) and Key Project of Education Department of Anhui (KJ2020A0008).}

\end{document}